\documentstyle[twocolumn,aps,pra,floats]{revtex}
\input epsf.sty
\begin{document}
\bibliographystyle{prsty}
\draft
\preprint{\today}
\title{ The Exchange-correlation Hole of the Si Atom, A Quantum Monte Carlo
Study}
\author{Antonio C. Cancio 
\footnote{Present Address: School of Physics, Georgia Tech, 
Atlanta GA 30332} and C. Y. Fong }
\address { Department of Physics, 
University of California, Davis, CA 95616 }
\author{ J. S. Nelson }
\address { Semiconductor Physics Division, Sandia National Laboratories,
Albuquerque, NM 87185-5800 }
\address{\rm (Submitted to Physical Review A)}
\address{\mbox{ }}
\address{\mbox{ }}
\address{\parbox{14cm}{\rm \mbox{ }\mbox{ }
We have studied the exchange-correlation hole $n_{xc}$ and 
pair correlation function in the valence 
shell of the Si atom in its $^3{\rm P}$ ground state, 
using highly accurate 
Slater-Jastrow wavefunctions and 
the Variational Monte Carlo method.  The exchange-correlation hole
shows a number of interesting features caused by the open
shell structure of Si, including a marked transition
from efficient to poor screening behavior as a test majority-spin 
electron is moved from
the center of a valence $p$~orbital onto the axis perpendicular to the occupied 
$p$~orbitals.  This behavior results from the dramatic difference in the
exchange hole in the two cases, which is partially compensated by a
corresponding anisotropy in the correlation hole.  
In addition we observe an anisotropic change in the spin density induced by 
Coulomb correlation, reducing 
the spatial overlap between the different spin-components of 
the density and contributing to the anisotropy of the correlation hole.  
The exclusion effect correlation, in which a
$3s^2\!\rightarrow\!3p^2$ substitution excitation is allowed along 
the unoccupied axis and prohibited along the other axes, was found to 
have a noticeable effect on the
correlation hole and is partially accountable for its anisotropy; 
however, it is inconsistent with the observed changes in spin density.
In contrast to the longer range features, 
we find that the ``on-top" correlation hole is well
described by linear density functional theory, for a large range of
local density and magnetization.  
}}
\address{\mbox{ }}
\address{\mbox{ }}
\address{\parbox{14cm}{\rm PACS numbers: 31.25.Eb, 31.10.+z,
71.15.-m, 02.70.Lq}}
\maketitle


\narrowtext

\section{Introduction}

An understanding of the exchange-correlation hole and related quantities
such as the pair correlation function and exchange-correlation energy is
an important factor in the systematic development of accurate
density functional theory (DFT) methods for quantum chemistry and
solid state physics~\cite{JG}.  
Much progress has been 
made in understanding the pair correlation function in the homogeneous electron
gas or jellium, using accurate numerical techniques~\cite{PickB,OrtizB} and 
analytic modeling~\cite{PW}.  
Although the closely related Coulomb hole has 
long been studied in atomic and molecular systems~\cite{Coulson,Sanders,CHoleRev},
the quantitative understanding of correlation holes
in inhomogeneous systems 
is far from complete.  In recent years there have been several efforts to 
fill the gaps, with a focus on characterizing properties closely
tied to the development of density functional theories, such as the
on-top hole~\cite{EPB,BPEontop} 
and system-averaged exchange and correlation holes~\cite{EPB,EBPmol}.

In much of the previous work on the exchange-correlation hole in
atoms and molecules, the focus has been on two electron 
systems and closed-shell atoms such as Be and Ne, 
with additional studies of more complicated molecules 
such as ${\rm N_2}$ and ${\rm H_2O}$~\cite{CHoleRev}.  
One class of systems that has received less attention is that
of open-shell atoms.  The valence shell of such an atom may consist only of
a few electrons about a radially symmetric core, but with a degenerate ground
state and the corresponding valence-shell structure, have interesting and
quite complex features in its exchange-correlation hole.  In particular,
in the case of less-than-half filling, these include the exclusion-effect 
correlations involving three or more electrons~\cite{Sinanoglu}.  
Modeling such systems requires proper attention both to 
orbital correlations best treated in a configuration
interaction (CI) context and to dynamic correlations 
more amenable to a density functional description. 
The application of density functional theory to open-shell systems
is an active area of research, with the question of the  
optimal treatment of degenerate ground states and of the formation
of such atoms into molecules giving rise to intriguing problems~\cite{Savin}. 


Recently, Variational Monte Carlo (VMC) methods 
combining highly accurate trial wavefunctions with Monte Carlo
methods for evaluating ground state energies and wavefunctions have been 
developed and
applied to atoms and molecules~\cite{Umrigar,SM,Mitas}.
The wavefunctions used in these calculations have the advantage of being
simple and compact, typically recovering 
$85\%$ or more of the correlation energy of atoms and the dissociation
energies of molecules with a few trial parameters.
Unlike configuration interaction expansions they can
describe with equal ease both ``nondynamic"
correlations and ``dynamic" correlations such as the short-range 
cusp condition~\cite{Kimball}.
These features make this method a natural candidate for studying electron 
correlations, and it has been employed in recent studies of the pair correlation
function in crystalline Si~\cite{FWL,Hood}.


In this paper we study the exchange-correlation hole and pair
correlation function of the valence
shell of the Si atom with the VMC method.  
A quantitative knowledge of correlations in Si is important in improving
DFT predictions for the cohesive energy, binding 
energies and surface characteristics for this and related technologically
important materials. 
The Si atom is in its own right a useful laboratory for
the study of electron correlations, 
with both local jellium-like and the nonlocal chemical properties
due to the open shell structure of the atom playing important roles
in determining the relevant physical and chemical properties.
The paper is organized as follows: Sec.~\ref{theory_nxc} provides 
theoretical background on the exchange-correlation hole, Sec.~\ref{theory}
gives details of the system studied and of the calculational method.
Results for the exchange-correlation hole are given in Sec.~\ref{results_nxc},
for the pair correlation function in Sec.~\ref{results_pcf} and
for correlation effects on the spin density in Sec.~\ref{results_spin}. 
We close with a summary of the results and conclusion in Sec.~\ref{conclusion}.


\section{The Exchange Correlation Hole}
\label{theory_nxc}
The exchange-correlation hole, including 
explicit spin dependence, is defined as:
\begin{equation}
   n_{xc}({\bf r}_0,\sigma_0; {\bf r},\sigma ) = 
     \frac{\displaystyle n^{(2)}({\bf r}_0,\sigma_0; {\bf r},\sigma)}
         {\displaystyle n({\bf r}_0,\sigma_0)} - n({\bf r},\sigma).
  \label{eqnxc}
\end{equation}
Physically it describes the change in density 
at ${\bf r}$ and for spin component $\sigma$ from
its mean value, $n({\bf r},\sigma)$, given the presence of another electron 
with spin $\sigma_0$
at position ${\bf r}_0$.  
The quantity $n^{(2)}({\bf r}_0,\sigma_0; {\bf r},\sigma)$
is the pair density, the expectation in the ground state of finding
a pair of electrons at two given coordinates.  It is defined
in terms of an expectation of the ground state as:
\begin{equation}
   n^{(2)}({\bf r}_0,\sigma_0; {\bf r},\sigma) = \sum_{i,j\neq i}
     \langle\delta({\bf r}_0 - {\bf r}_i) \delta_{\sigma_0\sigma_i}
     \delta({\bf r} - {\bf r}_j) \delta_{\sigma\sigma_j}\rangle.
  \label{eqnpair}
\end{equation}
A closely related quantity, the pair correlation 
function $g({\bf r}_0,\sigma_0,{\bf r},\sigma)$, 
is a measure of the pair density relative to that expected for uncorrelated
electrons with the same density distribution:
\begin{equation} 
   g({\bf r}_0,\sigma_0; {\bf r},\sigma) =
 \frac{\displaystyle n^{(2)}({\bf r}_0,\sigma_0; {\bf r},\sigma)}
      {\displaystyle n({\bf r}_0,\sigma_0) n({\bf r},\sigma)}.
  \label{eqpcf}
\end{equation}
The importance of the exchange-correlation hole to density functional 
theory lies in its connection~\cite{Exc2nxc,Levy} to the 
exchange-correlation energy $E_{xc}$:  
\begin{equation}
   E_{xc} = \frac{1}{2}\int\!d^3r\: n({\bf r}) \int\!d^3r^{\prime} 
   \int_0^1 \!d\lambda\: 
     \frac{\displaystyle n_{xc}({\bf r},{\bf r}^{\prime},\lambda)}
      {\displaystyle  |{\bf r} - {\bf r}^{\prime}|}.
   \label{eqexc}
\end{equation}
Here $n_{xc}({\bf r},{\bf r}^{\prime},\lambda)$ is the exchange-correlation 
hole, summed over spins, for the
system with scaled Coulomb interaction $\lambda e^2$, with an external potential
altered so that the density of the system remains unchanged.
The integration over coupling constant strength $\lambda$
accounts for the kinetic energy cost of correlating electrons, weakening
the strength of the integrated $n_{xc}$ with respect to its value at 
$\lambda\!=\!1$.
Although $E_{xc}$ is in principle determined from a knowledge of the single
particle density alone, this dependence is in general not easy to 
determine beyond the local density approximation.  
The exchange-correlation hole has a wealth of features which
may be used to test and improve theoretical models of the 
exchange-correlation energy. 
As a result, many 
attempts to systematically improve density functional theory have this
function as a starting point~\cite{JG,GJL,Becke,PBW-GGA}.

In this paper, we will discuss the full coupling-constant ($\lambda\!=\!1$)
case for $n_{xc}$.  Although the coupling-constant integrated quantity
is most directly connected to the density functional theory, the
full coupling-constant case is interesting in itself, as its average is 
an experimentally measurable expectation of the ground state~\cite{Thakkar}.
It is also an essential ingredient in modern hybrid methods which
combine elements of density functional theories and 
conventional Hartree-Fock methods~\cite{Becke}.
The quantity $n_{xc}$ is often analyzed by a decomposition 
into an exchange component, $n_x$, corresponding to the $\lambda=0$ or 
noninteracting system, which describes the correlations between particles
arising from the Pauli exclusion principle, 
and a correlation component, $n_c$, determined by taking the
difference between the fully interacting and noninteracting cases, 
which describes the additional correlation due to 
the Coulomb interaction between electrons~\cite{Coulomb}.  
We study the explicit spin decomposition of $n_{xc}$. 
This choice 
is useful for understanding the correlation response of
an open-shell atom for which the ground state has nontrivial differences in
its spin components.   
Spin decomposition falls roughly along the lines of exchange and 
correlation: exchange affects only particles with the same 
spin, and Coulomb correlations, though also present in the same-spin case, 
are most noticeable in the opposite-spin channel.

Some initial insight into the nature of the exchange-correlation hole in 
atoms and other finite systems
can be obtained by considering limiting cases.  The exchange-correlation
hole about a reference electron in the homogeneous electron gas
is isotropic, that is, a function solely of the distance from the electron, 
and localized about the electron
with a radius determined by the average interelectron distance.
Consequently, in a system of slowly varying density, the 
the local density approximation (LDA) holds, in which the hole
is determined by the density (or each spin-component of the density
if these are different) at the location
of the reference electron and ``moves" with the position of the reference
electron.  
In atoms or molecules, important correlation effects often involve a pair 
excitation from the noninteracting ground-state into a finite number of
lowlying, perhaps nearly degenerate,
excited states.  In this case, the resulting correlation hole is dependent
on the shape of the ground-state orbitals vacated and the 
excited-state orbitals occupied and is
largely unsensitive to the position of the
reference electron.  In real systems, the exchange-correlation hole
will contain aspects of both limiting cases, with those that ``move" with 
the position of the electron termed ``dynamic" correlations and the orbital
correlations insensitive to electron position termed ``nondynamic".  In
open-shell atoms both play important roles.


\section{Calculation Approach}
\label{theory}
\subsection{Correlated description of the Si atom}

We focus on correlations in the valence shell of the
atom with the core electrons replaced by norm-conserving {\it ab initio}
nonlocal pseudopotentials derived from LDA calculations~\cite{HSC}.
In this case the valence shell of the atom is described by the
Hamiltonian
\begin{eqnarray}
\nonumber H & = & \sum_i \left( \frac{\nabla_i^2} {2m} + \sum_{l_i, m_i}
	    V_{l_i}( r_i ) |l_i m_i\!><\!l_i m_i| \right) \\
            & & + \sum_{i<j} \frac{e^2} {|{\bf r_i} - {\bf r_j}|}
\end{eqnarray}
where the sums are over the valence electrons.  $V_{l}$ is the nonlocal
pseudopotential and $|l_im_i\!><\!l_im_i|$ the single particle projection 
operator onto the state with total angular momentum $l_i$ and $z$-axis 
projection $m_i$.
The final term is the intravalence Coulomb interaction. 

The Si atom has a 
nine-fold degenerate $(3s^2 3p^2)$ $^3P$ ground state.  By maximizing the
spin projection, this state can be
represented by a single Slater determinant, consisting of a 
majority-spin component, here chosen to be spin up, 
with one $3s$ and two $3p$ orbitals, and a minority or down-spin component with 
one $3s$ electron.  There then remain two angular momentum projections that
lead to physically significant differences in $n_{xc}$. 
The $m_l\!=\!0$ projection has $p_x$ and $p_y$
orbitals and a ``pancake" like shape, while the 
$m_l\!=\!\pm1$ projections have a ``cigar" shape, with $p_0$ and 
$p_{\pm}$ orbitals.
This distinction does not play a role in determining $E_{xc}$ for an
atom, given the invariance of the energy to rotations of the atom.
The presence of a quantization axis in the formation of a Si bond
breaks this
invariance, so that the projection-specific behavior of $n_{xc}$
becomes important in determining accurate molecular binding energies. 
It thus should be useful as a test of density functional theory, which typically
overestimates binding energies by about 1~eV~\cite{JG,EBPmol}.
Results in this 
paper focus on the $m_l\!=\!0$ projection which provides a clear comparison
between the situation perpendicular to and parallel to the $p$ orbitals, though
calculations were done on the other projection for comparison.

To calculate the exchange-correlation hole variationally we start with a 
Slater-Jastrow trial wavefunction
\begin{equation}
     \psi = exp(-F)\prod_{\sigma} D_{\sigma}, 
     \label{eqhfpsi}
\end{equation}
with $D_{\sigma}$ being a
Slater determinant for spin component $\sigma$ and $F$ a Jastrow correlation
factor.  All single-particle orbitals are obtained from
the same local DFT program that determined our pseudopotentials.  
We use a Boys and Handy form~\cite{SM,Boys} for $F$,
which includes electron-electron, electron-nucleus and 
electron-electron-nucleus correlations expanded in a basis set of correlation
functions:  
\begin{equation}
 F = \sum_{l,m,n} c_{lmn} \sum_{i\neq j} (\bar{r}_{i}^l \bar{r}_{j}^m + 
   \bar{r}_{i}^m \bar{r}_{i}^l) \bar{r}_{ij}^n.
   \label{eqjastrow}
\end{equation}
The basis functions are $\bar{r}_{i} = br_i / (1 + br_i)$ and
$\bar{r}_{ij} = dr_{ij} / (1 + dr_{ij})$, where $r_{ij}$ is the 
distance between a pair of electrons, $r_i$ is the distance between
electron $i$ and the atom center.  The terms $b$, $d$ and $c_{lmn}$ are 
variational parameters. 
The lowest order $\bar{r}_{ij}$ term is set separately for opposite and
same-spin electron correlations to satisfy the short range electron-electron
cusp condition~\cite{Kimball} for each case.
Higher order terms treat longer range effects 
and are determined without distinguishing electron spin.
Electron-ion terms in the correlation function correct for the 
tendency of interelectron correlations to expand the volume of 
the atom and provide density-dependent corrections to the electron-electron
correlation~\cite{SM}.
Since the valence shell of Si is less than half-filled, orbital or 
nondynamic correlations may be important~\cite{Sinanoglu}.  These can be
incorporated with a multideterminant extension of the Slater-Jastrow 
wavefunction:
\begin{equation}
     \psi = exp(-F)\sum_{\alpha} \eta_{\alpha} 
	  \prod_{\sigma} D^{\alpha}_{\sigma}.
     \label{eqcipsi}
\end{equation}

\subsection{Method of Calculation}
%
The Variational Monte Carlo method~\cite{CepK} is used to calculate the 
ground-state energy, 
derivatives with respect to variational parameters,  
and other expectation values.  The heart of the method lies in the 
judicious statistical sampling of integration points to obtain an 
estimate of the many-body integrals 
involved in evaluating expectations of the Slater-Jastrow
trial wavefunction.  This estimate is limited in accuracy by 
statistical noise; however if the trial wavefunction
is a good approximation to the ground-state wavefunction this noise can be 
very easily managed with a relatively small number of
configurations~\cite{Umrigar}.
Optimized wavefunctions are obtained by minimizing the 
variance of the energy~\cite{Umrigar}.
With a trial wavefunction of 18 expansion terms including two
set by the same- and opposite- spin cusp conditions, up to $l+m+n=6$
in the basis function expansion, we obtain a value
of 3.8028(2)~a.u.\ for the ground-state valence-shell energy.  The
correlation energy, measured with respect to a noninteracting ground-state
energy of 3.7188~a.u., is 97$\%$ of the correlation energy obtained
from Green's function Monte Carlo and 95$\%$ of that obtained from CI 
using the same nonlocal pseudopotential~\cite{Mitas}.
A similar calculation starting from a two determinant reference 
point [Eq.~(\ref{eqcipsi})],
adding the $3p_z^23p_x3p_y$ excited state to the noninteracting ground-state
configuration, resulted in a modest improvement in energy to 3.8041(2)~a.u.\
or 96.6$\%$ of the valence-shell correlation energy with respect to CI.
%

To calculate correlation functions, we measure spin-decomposed single-particle 
densities $n({\bf r}, \sigma)$ and conditional densities 
$n({\bf r},\sigma|{\bf r}_0,\sigma_0)$.  The spin-dependent conditional 
density is defined as the ground-state density distribution 
as a function of spin $\sigma$ and position ${\bf r}$ of the
$N\!-\!1$ other particles given one with spin ${\sigma_0}$ fixed at ${\bf r_0}$.
The difference between the conditional and ``unrestricted" densities,
gives the spin-dependent exchange-correlation hole, Eq.~(\ref{eqnxc}),
\begin{equation}
     n_{xc}({\bf r_0},\sigma_0;{\bf r},\sigma) =
	   n({\bf r},\sigma|{\bf r}_0,\sigma_0) - n({\bf r},\sigma).
     \label{eqcdens}
\end{equation}
Separate calculations are done to measure the density and the conditional
density for various values of ${\bf r_0}$ and $\sigma_0$.  
These expectations are first
calculated exactly for the Slater-determinant wavefunction (setting
$F\!=\!0$ in our trial wavefunction).  Then, the difference between the
expectations obtained with the Slater determinant and the fully interacting 
wavefunctions is measured statistically using Monte Carlo sampling and the 
method of correlated estimates~\cite{CepK}. 
This technique is an efficient means to estimate statistically
the change in the expectation of an observable under a small perturbation of 
the Hamiltonian or of the variational parameters of the wavefunction, 
taking advantage of the high degree of correlation
between the two expectations to reduce noise in the
difference of their statistical estimates.  
In the present case, the correlation hole, which describes the 
difference between the correlated and Slater-determinant exchange-correlation
holes, is fairly small compared to the exchange-correlation hole.  It is thus
a reasonable assumption that the correlated estimation approach should
improve the sampling efficiency for this quantity.
In practice, the procedure was observed to reduce statistical
errors in our data by a factor of 5, and roughly $10^5$ random samples
of the wavefunction sufficed to obtain expectation values.


The expectations for density and conditional density were expanded in a plane 
wave basis, taking the average of $\sum_i exp( -i{\bf G}\cdot {\bf r}_i )$ 
for a set of plane waves up to a 32 Ry cutoff,
on a supercell 18 Bohr radii ($a_B$) in length. 
The coefficients were symmetrized and fast-Fourier transformed to obtain
real-space densities.
This expansion provides smooth profiles despite the statistical noise
of the sampling.  On the other hand the short-wavelength cutoff 
causes an unrealistic rounding off of the correlation hole cusp at
short interparticle distances, and spurious oscillations at low densities. 
At present this cutoff is the largest error in our calculation.  In the
case of the exchange-correlation hole this error is notable mostly  
in the cusp region.  A more complete discussion of errors is presented 
in Sec.~\ref{results_pcf} in regard to the pair correlation function which 
is more sensitive to the cutoff error than the exchange-correlation hole.

\section{Results for the Exchange-Correlation Hole}
\label{results_nxc}
The exchange hole in DFT is obtained by evaluating $n_{xc}$ from 
Eq.~(\ref{eqnxc})
for the Slater determinant wavefunction that minimizes the energy in the 
noninteracting ($\lambda\!=\!0$) limit: with the Coulomb interaction replaced 
by a single-particle potential that reproduces the true ground-state 
density~\cite{Levy}.
In practice the Slater determinant of local DFT orbitals used in our
calculations produces a density that differs from our VMC density
by a few percent.  In terms of these orbitals, the exchange hole is:
\begin{equation}
   n_{x}({\bf r}_0,\sigma_0; {\bf r},\sigma) = 
     \frac{ - \left| \sum_{\alpha=1}^{N_\sigma} 
            \psi^*_{\alpha}({\bf r}_0,\sigma) \psi_\alpha({\bf r},\sigma)
            \right|^2 \delta_{\sigma \sigma_0} }
         {\displaystyle n({\bf r}_0,\sigma_0)}
\end{equation}
where the sum runs over all the occupied single-particle orbitals $\psi_\alpha$
for the spin component $\sigma$.  This expression, as a function of
${\bf r}$ for fixed ${\bf r}_0$ ({\it i.e.}, interpreting the hole
as the change in density of the system at ${\bf r}$ given a particle observed
at ${\bf r_0}$) has the form
of minus the probability density of a hybrid atomic orbital.  That is, it
describes 
a normalized linear combination of single-particle orbitals with coefficients 
$c_{\alpha} = \psi^*_\alpha({\bf r}_0,\sigma_0) / \sqrt{n({\bf r}_0,\sigma_0)}$.
This choice of $c_{\alpha}$ for each orbital $\psi_{\alpha}$
represents 
the unique linear combination of orbitals that maximizes
the probability for an electron to be observed at 
${\bf r_0}$ and spin $\sigma_0$ (conversely giving zero likelihood for 
any other electron to be observed at that point.)
Finally the integral over ${\bf r}$ of the 
exchange hole is $-1$, as it measures the integrated difference in 
density between the $(N\!-\!1)$-electron system given one electron fixed 
at ${\bf r_0}$ and the full $N$-electron system.

\begin{figure*}[t]
\epsfysize=5.2cm
\epsfbox{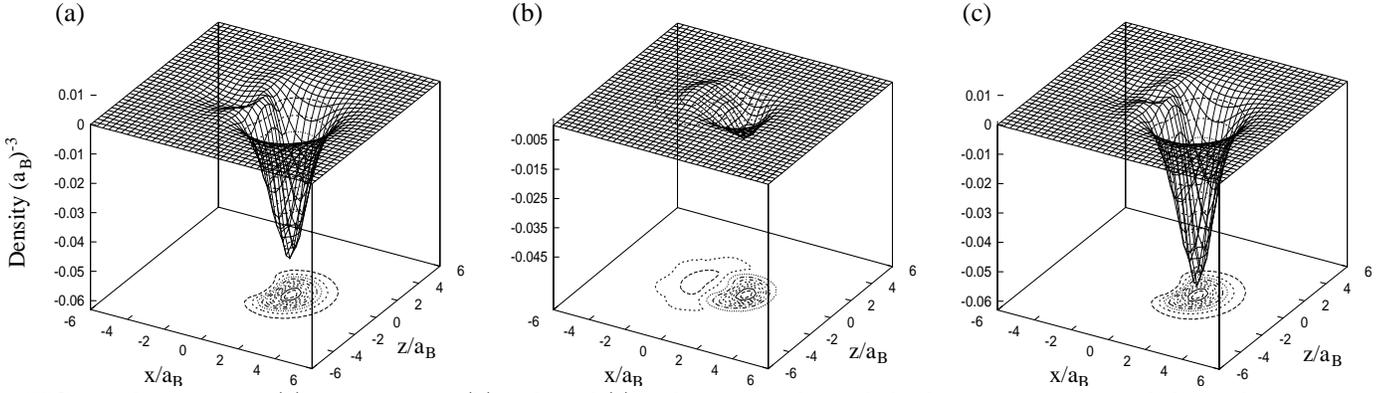}
\widetext
\caption
{The same spin (a), opposite spin (b) and total (c) exchange-correlation hole
about a spin-up particle located at 1.4~$a_B$ from the atom center on the
$x$~axis (parallel to the $p$ orbitals) for the ground state of the Si atom 
in the $L_z\!=\!0$ projection.  
The surface plot shows the change in density along a plane cutting
through the origin along the $x$ and $z$ axes.  }
\label{atomfig1}
\narrowtext
\end{figure*}

\begin{figure*}[t]
\epsfysize=5.2cm
\epsfbox{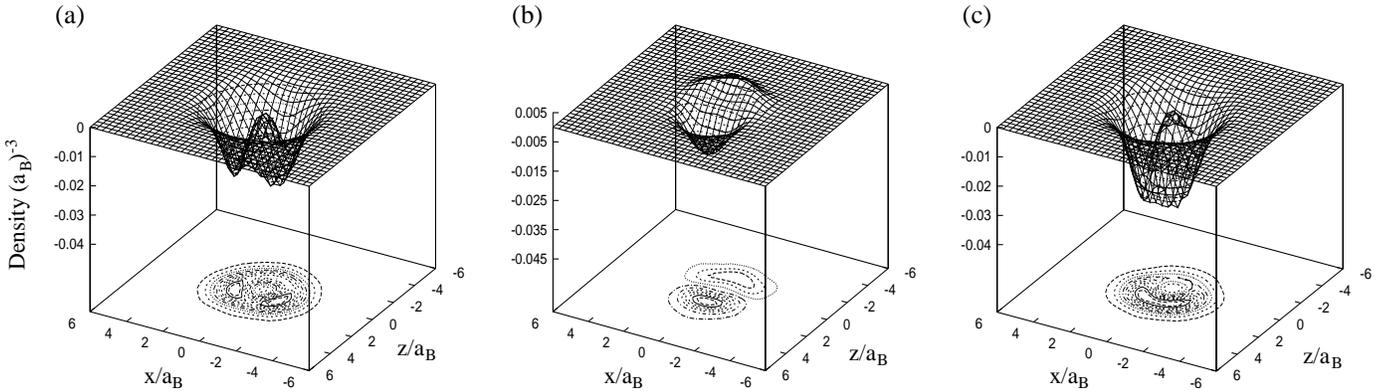}
\widetext
\caption
{The same spin (a), opposite spin (b) and total (c) exchange-correlation hole
about a spin-up particle located at 1.4~$a_B$ from the atom center on the
$z$~axis (perpendicular to the $p$ orbitals).  
Cut through atom as in Fig. 1.} 
\label{atomfig2}
\narrowtext
\end{figure*}

For a spin-up particle on the $z$~axis in the
$L_z\!=\!0$ projection, this exchange-hole orbital
is a $3s$ state since the occupied $p$ states ($p_x$,$p_y$) are 
in the $x$-$y$ plane.  The exchange
hole is completely insensitive to the electron position on this axis, given
only one possible orbital from which to construct it.  The situation along
this axis thus corresponds to an extreme departure from the ``dynamic" picture
of the exchange hole derived from the homogeneous electron gas.
For a particle on the $x$~axis, the hole is a combination of $s$ and $p_x$
orbitals.  We find that the occupation probability $|c_p|^2$ of the $p$ orbital
to be between $60\%$
and $75\%$ in the region $(0.5~a_B\! <\! x\! < \!2.5~a_B)$ of peak density 
along the
$x$~axis; this is roughly equivalent to that of a $sp^2$ hybrid orbital 
(which is two-thirds $p$) oriented
in the direction of the fixed particle.  The slight variation in the 
$c_p$ acts to keep the hole more or less centered on the reference electron
in the region of peak density on the $x$-$y$ plane.
The exchange hole 
as a result is more sensitive to the exact location of particles in
this plane, and therefore more efficient in screening them, than along
the $z$~axis, leading to significant differences in the correlation 
holes in the two cases as well.  
This transition from
``nondynamic" screening along the axis perpendicular to the occupied 
$p$~orbitals to a more ``dynamic" screening in the plane occupied by them, 
particularly as it occurs at peak density in the valence shell,
constitutes a major difference between the $n_{xc}$ of Si and that of 
closed-shell atoms.

In Figs.~\ref{atomfig1} and~\ref{atomfig2} we plot the exchange-correlation
hole about a spin-up electron fixed at the point of
peak density parallel to and
perpendicular to the two $3p$ orbitals of the $m_l=0$ (pancake) projection
of the Si atom.  The single-configuration Slater-Jastrow 
wavefunction [Eq.~(\ref{eqhfpsi})] is used.
Each plot shows the response to this electron in the $x$-$z$ plane,
that is, the plane cutting through the center of the
atom at the plot origin, with one axis ($x$) parallel to and one ($z$) 
perpendicular to the occupied $p$ orbitals.
The hole is split into same spin~(a), opposite spin~(b) and total~(c) response.
The comparison between these two situations shows the dramatic anisotropy
in $n_{xc}$ reflecting that of the exchange hole.

For a spin-up electron placed on the $x$~axis, Fig.~\ref{atomfig1}, 
the exchange-correlation hole is dominated by the $sp^2$-like 
exchange hole.
The additional effects of Coulomb correlation
on the same-spin channel are hard to detect, 
while the opposite-spin correlation hole 
is small (contributing $14\%$ of the total on-top hole, or value of
the hole at zero interparticle separation.)  It is largely 
confined to a narrow region about the electron, indicating that the
$sp^2$ hybrid hole screens the electron efficiently.  

Fig.~\ref{atomfig2} shows the exchange-correlation hole of a spin-up electron 
on the $z$~axis, perpendicular to the two $p$ orbitals.  
With the addition of correlation, the 
same-spin hole (a) loses the rotational symmetry of the $3s$ state that
characterizes the exchange hole, with 
the polarization of the two $3p$ orbitals creating a double valley on 
either side of the reference electron.
The opposite-spin hole (b) shows the polarization 
of the $3s$ spin-down orbital, with a well centered
about the fixed electron and a strong dipole response at longer range.
%

The total $z$-axis exchange-correlation hole, (c), 
shows a smooth interpolation of the two spin
contributions leading to a large crescent-shaped area near the electron
from which the other electrons are repelled.  The
correlation hole contributes considerably to the total
exchange-correlation hole,
with up to $40\%$ of the total on-top hole due to correlation.
Although the $3s$-orbital exchange hole is highly nonlocal and does not
efficiently screen the electron, the correlation
contribution goes a long way to make the total hole more local.  

\begin{figure}[ht]
\epsfysize=10.0cm
\epsfbox{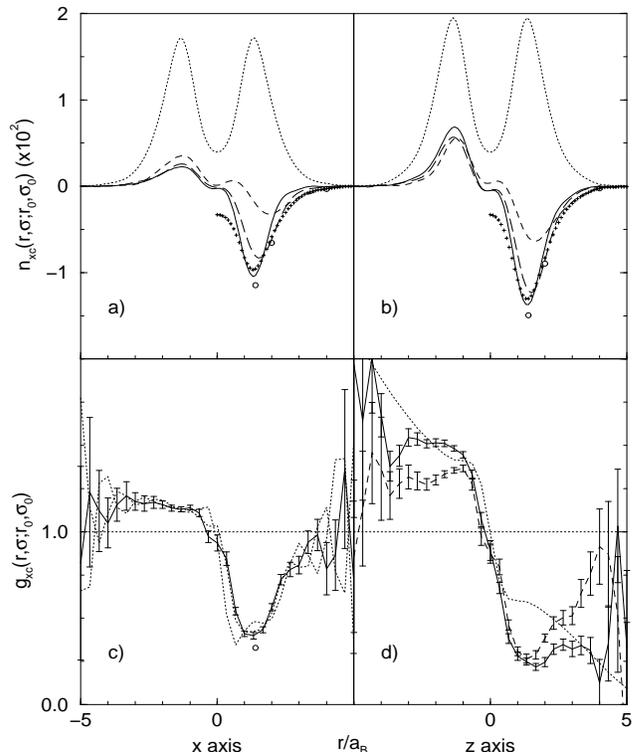}
\caption
{Spin-down response to a spin-up electron:  
(a) the correlation hole 
along the $x$~axis for a spin-up electron fixed on the $x$~axis
at positions $1.4$ (solid line), $2.0$ (long-dashed) and 
4.0~$a_B$ (dashed line) from the atom origin.
(b) the correlation hole along the $z$~axis for an electron
fixed at corresponding positions on the $z$~axis.  
(c) the pair correlation function for the 1.4~$a_B$ case of (a), for energy 
cutoff of 20, 28 (both dotted lines) and 32~Ry (solid line).
(d) the pair correlation function for the 1.4~$a_B$
case of (b) for three trial wavefunctions: 
$\psi_{S-J}$ (dashed line), $\psi_{CI}$ (dotted), $\psi_{CI-J}$ (solid line).
In (a) and (b), the mean spin-down density is plotted 
for comparison as a dotted line.  Additionally, the 
circles show the value for the ``on-top" hole for each
case plotted and the crosses, the LSD on-top hole for various points along 
each axis.
Error bars show the standard deviation of the VMC expectation 
measured over several statistically independent runs.}
\label{atomfig3}
\end{figure}

In addition to the obvious differences in $n_{xc}$ along each axis
due to the differences in the exchange hole, Figs.~\ref{atomfig1} 
and~\ref{atomfig2} reveal subtle differences in the opposite-spin hole. 
The extent of the orientational anisotropy in the opposite-spin hole
can be better visualized by plotting the hole along the
$x$~axis for the up-spin electron fixed on the $x$~axis and along the $z$~axis
for the electron fixed on the $z$~axis, cutting through the minima 
and maxima of the contour plots Figs.~\ref{atomfig1}(b) and \ref{atomfig2}(b).  
These are represented as solid lines in Figs.~\ref{atomfig3}(a) and~(b).  
The most notable difference between the two cases is the height of the
peak on the side of the atom opposite 
the reference electron, which is three times as large along the $z$~axis~(b)
as on the $x$~axis~(a).  In addition the minimum is slightly deeper for 
the $z$-axis case.  

Additionally, in Figs.~\ref{atomfig3}(a) and (b) we show trends in the 
opposite-spin correlation hole as one gradually removes the reference electron
from the atom.
In addition to the 1.4~$a_B$ case discussed above, we place
a spin-up electron on the $x$~axis~(a)
and the $z$~axis~(b), at a distance of 2.0 and 
4.0~$a_B$ from atom center, and plot the on-axis response of the spin-down
electron as long- and short-dashed lines respectively.  
These three reference radii correspond to placing an electron at the 
peak valence density along either axis, at the average radius from the atom,
and at a low density point outside the atom respectively.  
As the electron is moved to lower densities, the 
shape of the minimum slowly gets wider and shallower, consistent
with trends in the homogeneous electron gas.  The position of the hole 
minimum stays near the atom, and thus increasingly
more off center with respect to the electron.  This is expected: the correlation
hole, measuring the change in density in the presence of an electron 
at some reference point, can have an absolute value no greater than the 
density itself.
Along the $x$~axis, this trend to a shallow off-centered hole is correlated 
with a gradual increase of peak height on opposite side of atom.  
For the $z$-axis 
case the peak height remains roughly constant as the electron is removed.

It is interesting to compare these results to recent analytic studies 
of the asymptotic limit of the
exchange-correlation hole of Be and Ne~\cite{EPB,EBPasymp}.
In the limit of an electron far removed from the
atom, $n_{xc}$ measures the collapse of the remaining
$N\!-\!1$ electrons to an eigenstate of the positive ion.  The correlation
component of $n_{xc}$ measures the change in electron density due to the
reduced Coulomb reduction between the remaining electrons.  
At intermediate distances, a few atomic radii from the atom center, $n_{xc}$
also shows a dipole polarization of the atom, 
or a reduction of electron density from
the side of the atom nearer the reference electron to the farther side, more so 
for the more polarizable Be than for Ne.
The collapse to the ion in the asymptotic limit results in a correlation hole 
peaked in the center with a minimum
around the edge of the atom.  This may possibly be seen in our data 
for the 4.0~$a_B$ $x$-axis case of Fig.~\ref{atomfig3}(a), 
where the correlation-hole 
profile develops a peak on both sides of the ion core.  
The correlation hole along both axes shows evidence of a dipole 
polarization of the valence shell when the reference electron is moved 
outside the atom, 
with a larger response on the ``Be-like" $z$~axis than along the 
``Ne-like" $x$~axis.
The contrast between Si and closed-shell atoms is that this polarization
of the valence shell does not die out when a reference electron on the $z$
axis is moved into the center of the valence shell, if anything becoming more 
pronounced, while the polarization is damped out along the $x$~axis under
the same circumstances.  An added complexity for Si is that the
final state of the system after removing an up-spin electron in different
directions corresponds to different atomic configurations: 
$3s3p^2$ for the $z$-axis case (removal
of the up-spin $3s$ orbital) but $3s^23p$ for the $x$~axis.  Thus significant
differences between the correlation and exchange holes induced by electrons
on different axes remain even for electrons asymptotically far from the atom.

We have also studied the local features of the exchange-correlation holes 
plotted, especially the opposite-spin on-top correlation hole, 
$n_{xc}({\bf r}, \sigma; {\bf r}, -\!\sigma)$.
This measures the reduction in the opposite-spin electron density 
at the location of a given electron.  As the correlation hole is deepest when
the Coulomb repulsion is largest, the on-top hole also 
represents the lower
bound for the correlation hole at any position in the atom~\cite{ontop}.
In previous papers, it was observed that this feature was 
well described in several small atoms~\cite{EPB,BPEontop} and in 
Hooke's atom~\cite{PSB} by a local spin-density (LSD) ansatz, using 
the on-top hole of the homogeneous electron gas corresponding
to the density and magnetization at the location of the electron pair:
\begin{equation}
     n^{LSD}_{xc}({\bf r}, {\bf r}) = n({\bf r}) 
	   ( g^{heg}( 0, n({\bf r}), \zeta({\bf r}) ) - 1 ).
\end{equation}
Here $g^{heg}(u,n,\zeta)$ is the pair correlation function of the homogeneous
electron gas at density $n$, magnetization 
$\zeta = (n_{\uparrow} - n_{\downarrow})/n$, and interparticle separation $u$.
In this sense, it represents a 
link between the homogeneous electron gas and dynamic correlations
in many inhomogeneous systems~\cite{PEBS}.

In Figs.~\ref{atomfig3}(a) and (b) we plot the LSD on-top hole, using
homogeneous electron gas values from Ref.~\cite{Yasuhara}, at each 
point along the $x$ and $z$ axes (crosses). 
These are compared to on-top holes of the Slater-Jastrow wavefunction 
[Eq.~(\ref{eqhfpsi})] evaluated specifically at the location of the 
electron for the correlation holes shown 
(circles at 1.4, 2.0, and 4.0~$a_B$).
These values for the Slater-Jastrow on-top hole are calculated 
directly without recourse to a plane-wave expansion,
using VMC with a constrained pair random walk~\cite{CanCYF},
and are exact within a negligible statistical error.
The finite energy cutoffs employed for the rest of the data tend to lose
sharp features of the correlation hole, in particular that of the cusp
condition at zero interparticle separation.  The resulting disagreement with
the exact (for the trial wavefunction used) values is at worst $20\%$ and
usually better.
The LSD model agrees fairly well with the VMC data, within
$10-20\%$ for most of the points calculated here,
faithfully following the observed trends with respect to electron position.
It provides a fairly reasonable
lower limit for the correlation hole at all points in the atom.
A calculation of the on-top hole about the spin-down electron was 
also done, with similar agreement with theory.  


It is interesting to consider in further detail 
the anisotropy in the opposite-spin 
contribution to $n_{xc}$ at peak density.  In the cases shown in 
Figs.~\ref{atomfig1}-\ref{atomfig3}, 
it can be interpreted as the measure of
the polarization of the down-spin electron density in the 
presence of an up-spin electron.  At first glance, one might guess that this
down-spin polarization should roughly
be invariant with respect to the angular orientation of the up-spin electron
since one is measuring the response of the down-spin $s$-orbital 
to the angle-independent Coulomb interaction.  More precisely, one measures
the correlations induced by the optimized Jastrow factor 
which depend only on interelectron and electron-nucleus
distances.  
The difference in the opposite-spin holes 
Figs.~\ref{atomfig1}(b) and 
\ref{atomfig2}(b), is thus a direct reflection of the influence of the 
determinantal structure of the three up-spin electrons on the correlation
response of the down-spin one.

In the LDA, the influence of the environment on interparticle correlation
is modeled by the variation with local density of the correlation 
hole, taken to be that of the homogeneous electron gas~\cite{OrtizB,PW} 
at the local density.
Typically the largest absolute and relative effects in this model
are felt at the on-top hole.  At this point it seems to 
describe the observed behavior quite well, with the variation in density
going from the maximum of the $p$ orbital on the $x$~axis to the open
$z$~axis accounting for the difference in the peak density on-top holes.
However, the large discrepancy 
between the maxima of the peak density $z$-axis and $x$-axis hole,
the solid lines in Fig.~\ref{atomfig3}a and~\ref{atomfig3}b, 
as well as their relatively large size, is inconsistent with this model. 
Neither does it seem easily explained by local gradient corrections, 
as the two points are located at a density maximum and a saddle point
respectively, where any correction
would be only second order in the gradient and therefore quite small.


Within the context of a variational wavefunction calculation, a
mechanism that can introduce an orientation dependence into the
correlation hole of an open shell atom
is the three-electron ``exclusion effect" or Fermi 
correlation~\cite{Sinanoglu} in configuration-interaction theory, which 
should be observable for second row atoms with less
than half filling in the $3p$ valence shell.  For Si in the
$L_z\!=\!0$ projection considered here, there is 
a ``nondynamic" contribution to the correlation between electrons 
along the $z$~axis due to a $3s^2$ to $3p_z^2$ excitation that is forbidden
for the corresponding $x$- or $y$-axis analogs by Pauli exclusion.
This contribution to the correlation hole about an
up-spin electron on the $z$ or ``unoccupied" axis, and the absence of 
the same about one on ``occupied" axes leads to a
difference in the holes observed with reference points on these axes.
Since this particular configuration is within the $3p$ valence shell, 
and thus fairly close to the ground state energy, it
is natural to expect it to play a significant role in the observed
difference in response.

What is striking here is that the observed orientational dependence of 
$n_{xc}$ is obtained with a single Slater-determinant
configuration modified by a Jastrow factor that depends only on 
interparticle distance and the distance of particles from the core.  
There are no explicit orbital correlations in this trial wavefunction 
[Eq.~(\ref{eqhfpsi})].
The Slater-Jastrow wavefunction, coupling a Jastrow factor 
with a determinant formed from an incomplete set of valence orbitals,
in principle could have a nonzero overlap with the $3p_z^23p_x3p_y$ 
configuration allowed in the normal exclusion-effect model, and no overlap
with the excluded excitations.
However, this would be only one of
an infinite number of configurations implicitly included in the Slater-Jastrow 
wavefunction, tied together by a ``dynamic" correlation factor that
gives no particular variational weight to any one configuration.  Thus
it is unlikely that it would fully account for the observed differences
in the response along the occupied and open axes~\cite{Node}. 

On the other hand, a more general argument involving screening can be 
invoked to explain the orientational anisotropy in our results.
If the up-spin electron is placed on the $z$-axis, its exchange hole 
is not centered on the electron position, as the electron occupies
a $3s$ orbital with an isotropic spatial extension 
throughout the atom's valence shell.  The net effect of the electron 
plus its exchange hole has the nature of a dipole field along the
$z$~axis of the atom,
with an imperfectly screened negative charge on the side of the atom nearest
the electron and a positively charged
region due to the exchange hole on the other side.
The large peak in the down spin electron's response to the up-spin electron
can be viewed then as a dipole induced by the dipole field
formed by the imperfect exchange screening of the up-spin electron.
In the $x$-axis case, the exchange
hole is centered and localized on the electron; the other two electrons
are in $sp^2$ orbitals blocking out the rest of the valence shell.  
Therefore the up-spin electron is efficiently screened by its exchange
hole, and in addition, there is 
no ``easy" direction for redistributing the down-spin electron density.
In such a picture the $3s^2 \rightarrow 3p^2$ excitation should play an 
important but not exclusive role in the overall dipole response 
along the $z$~axis.
This effect is observable in a Slater-Jastrow wavefunction because the 
Jastrow factor includes correlations simultaneously between all particle 
pairs.  
Thus, the correlation response of the down-spin electron to a fixed up-spin
electron involves not only their mutual Coulomb repulsion but that of the
other up-spin electrons as well.  The inhomogeneous distribution 
of these electrons about the fixed electron combined with the effect of the
external potential in essence create a Coulomb interaction with the exchange
hole.  

\section{Results for the Pair Correlation Function}
\label{results_pcf}
\subsection{Single Configuration Wavefunction}


In a system such as an atom, the pair correlation function 
$g({\bf r_0},\sigma_0; {\bf r}, \sigma)$
is a valuable tool
because, given the large variation in the density over the length-scale of the
exchange-correlation hole, the shape of the hole is to a large extent
determined by the variation in density.  The pair correlation function
can distinguish the intrinsic effects of correlation 
by eliminating any features that are simply proportional to the density.

\begin{figure*}[t]
\epsfbox{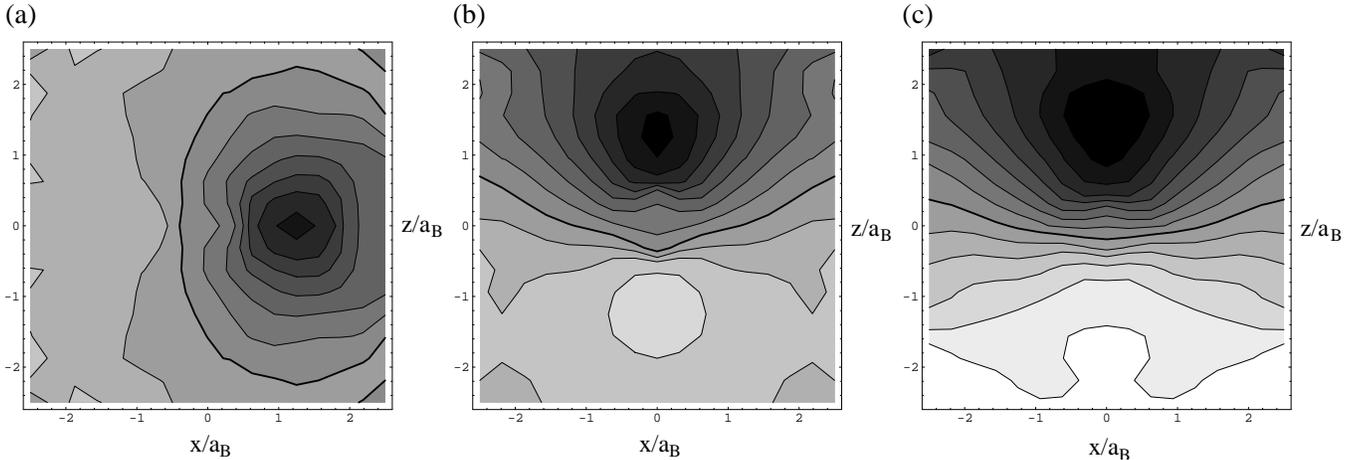}
\widetext
\caption
{The opposite-spin pair correlation function 
$g(\bf{r}_0,\uparrow; {\bf r}, \downarrow)$ 
as a function of spin-down electron position in the $x$-$z$ plane for 
a spin-up particle fixed at (a) 1.4~$a_B$ from the atom center on the
$x$~axis, (b) 1.4~$a_B$ from atom center on the $z$~axis, (c) the same as (b),
with nondynamic correlations from the $3s^2$ to $3p^2_z$ double
substitution included.  The contour increment is 0.1 with the
contour for $g\!=\!1$ represented by a thicker line; the regions
between are shaded darker for lower values of $g$.}
\label{atomfig4}
\narrowtext
\end{figure*}

In Fig.~\ref{atomfig4} we plot the pair correlation function (PCF)
for several of the cases considered previously.  Specifically, we show 
$g({\bf r_0},\uparrow; {\bf r}, \downarrow)$ fixing a spin-up
particle at the peak in the valence density, 1.4~$a_B$ on the 
$x$~axis (a) and the $z$~axis (b) and
plotting the variation with respect to the spin-down electron 
in the $x$-$z$ plane.  These results are
obtained by dividing $n_{xc}$ as plotted in
Figs.~\ref{atomfig1}(b) and \ref{atomfig2}(b)
respectively by the spin-down density, with a constant shift of 1 due to
the differing conventions in their definitions [Eqs.~(\ref{eqnxc}) 
and (\ref{eqpcf})].  
The value $g\!=\!1$ corresponding
to $n_{xc}\!=\!0$ (shown as a thick contour in Fig.~\ref{atomfig4}) 
denotes the boundary between the region
in which the spin-down density has been reduced ($g\!<\!1$)
and that in which it has been enhanced ($g\!>\!1$) in the presence of the 
spin-up reference electron.  
As described in the next 
section the PCF\ can be significantly affected by statistical error at
low density, so that the plot range is restricted to higher density regions
where the statistical error is less than half a contour increment in
either direction.  

The contours of the pair correlation function in the vicinity
of the spin-up electron on the $x$~axis are nearly circular and
centered on the electron,
showing the spin-down electron repelled from it without significant
directional bias.  The $g\!=\!1$ contour is slightly oblong in shape, 
indicating that at longer distances there is a slight bias towards
shifting electron density to the high density areas on the
far side of the atom ($x < 0$).  
A similar plot cutting through the $x$-$y$ plane (ie, cutting through both 
the $p$ orbitals along a plane perpendicular to the one shown) 
shows similar results except that the $g\!=\!1$ contour is
more circular in shape.  
This picture is fairly consistent with that of a homogeneous electron
gas PCF\ which depends only on the distance between particles, and indicates
that the underlying physical processes are quite similar: high energy
``dynamic" correlations that are quickly screened out over the radius of the
exchange hole. 

In comparison, the PCF\ in Fig.~\ref{atomfig4}(b), 
showing the
response to the spin-up electron on the $z$~axis shows a much greater
departure from isotropy.  The function has 
a noticeable dipole component with respect to the atom center, with a 
peak in $g$ directly opposite the minimum on the $z$~axis.  The $g\!=\!1$
contour has a much more shallow curvature than that of 
Fig.~\ref{atomfig4}(a), so that it is no longer centered
on the spin-up electron, but rather coincides roughly with 
the $z\!=\!0$ plane in the high density region of the atom. 
(For the $3s^23p_x3p_y$ configuration of Si, the PCF\
with a reference electron on the $z$~axis
is rotationally invariant about the $z$~axis so that the contours shown
in the plot can be extended to surfaces of rotation in three dimensions.)
In the near vicinity of the fixed electron, the pair
correlation function is no longer a function of the
distance between the two coordinates ${\bf r}$ and ${\bf r_0}$ alone, 
but has taken on a noticeable angular dependence as well.  
Thus the slope of the
well near 1.4~$a_B$ on the $z$~axis is steeper towards the center of the
atom, shallower towards the outer edge of the atom.
Although such a shape could in 
principle be accounted for in a region where the gradient of the density
is large, using a gradient expansion of the homogeneous
electron gas~\cite{Langreth}, it cannot be explained along these lines
in the current situation since it occurs at local saddle point in the 
density (the peak of the density along the $z$~axis).
These features in the PCF\ confirm the existence
of a genuine directional anisotropy in the response
to an electron located perpendicular to the occupied $p$ orbitals,
and suggest that the explanation lies in a nonlocal 
mechanism such as the response to a poorly localized $3s$ exchange hole.

Along this line, it is instructive to consider the exchange-correlation
hole about the lone down-spin electron.  In this case the exchange hole
is of necessity due to the $3s$ orbital regardless of the electron 
position, and one can expect significant departures from a compact, 
isotropic PCF.  In the case of a down-spin electron on the $z$~axis, we 
observe the PCF that is very close to that about an up-spin electron, 
with perhaps a slightly larger dipole component.  In both cases, 
the PCF indicates the presence of a significant dipole response that
compensates for the poorly screening $3s$ exchange hole.
The PCF for a down-spin electron on the $x>0$~axis has a quite complicated
pattern outside the on-top hole region -- neither roughly isotropic like
Fig.~\ref{atomfig4}(a) or with simple dipole anisotropy like 
Fig.~\ref{atomfig4}(b).  There are several unconnected regions where the 
electron density is enhanced: surrounding the electron for $x>0$ and 
focused on the $x$~axis opposite the electron for $x<0$.
In this case the exchange hole should not screen the electron
efficiently, but there is no lowlying $3p$ orbital with which
to construct a response, so that there is a possibility that the
contributions of higher-order orbitals might be observable. 


\subsection{Two Configuration Wavefunction}

Since we have observed a
significant change in shape of the opposite-spin PCF\ 
with respect the angular orientation of the reference electron,
using a single-determinant Slater-Jastrow wavefunction,
it is interesting to include explicitly the $3s^2 \rightarrow 3p_z^2$ 
excitation that is predicted in CI theory to play
a prominent role in producing such a behavior in the Si atom.
We consider the multideterminant wavefunction $\psi_{CI}$ that
consists of a $3p_z^23p_x3p_y$ excited state configuration as well as the
noninteracting ground state.  This wavefunction can be treated alone
or used as a starting point for adding further correlations via the 
Jastrow factor [Eq.~(\ref{eqcipsi})]; 
in either case the mixing amplitude $\eta_1$ for the excited state
is variationally optimized, along with the Jastrow
parameters for the latter case.  We obtain a mixing coefficient $\eta_1$ of
0.132 and total energy 3.7263(18)~a.u.\ in the former case and 0.056 and
3.8041(2)~a.u.\ in the latter.

The nondynamic correlation modeled by the two-configuration wavefunction 
alone, $\psi_{CI}$,  is to first order in $\eta_1$ a 
dipole-dipole correlation:
\begin{equation}
g_{CI}({\bf r_0},\uparrow;{\bf r},\downarrow) = 1 + 2\eta_1 
   \frac{z_0}{r_0}\frac{z}{r} 
   \frac{R_p(r_0)}{R_s(r_0)}\frac{R_p(r)}{R_s(r)}
   \label{eqgci}
\end{equation}
where $R_p$ and $R_s$ are the radial $3s$ and $3p$ orbitals.
The signature of this PCF\ in a contour plot with the cut through the
$x$-$z$ plane and $r_0$ fixed on the $z$~axis is a series 
of roughly straight lines arranged antisymmetrically about the $z\!=\!0$ plane, 
with the $g\!=\!1$ contour at $z\!=\!0$.
The shape of the function is independent of the position of the 
fixed particle, with the only change being in the overall amplitude.
For one electron fixed on the $x$~axis, on the node of the $p_z$ orbital, the
PCF\ is to first order zero.

In Fig.~\ref{atomfig4}(c) we show a contour plot for the opposite-spin 
PCF\ of the combined CI plus Jastrow
wavefunction, $\psi_{CI-J}$, for the same plot parameters as 
Fig.~\ref{atomfig4}(b):
with the up-spin electron fixed at 1.4~$a_B$ on the $z$~axis.
In this case the occupation probability of the excited-state 
Slater-Jastrow wavefunction is very small, 0.3$\%$, 
and the change in the total ground-state energy from the single-determinant 
Slater-Jastrow wavefunction is correspondingly small.  
On the other hand the explicit addition of the nondynamic
correlation increases the correlation energy by 1.6$\%$ and
has a significant impact on the shape of the hole.  
The PCF has a clear dipole-dipole signature,
exaggerating the spatial anisotropy already visible in the one determinant
case.  The $g\!=\!1$ contour is closer to the $z\!=\!0$~axis, 
and the short-range 
well about the fixed particle, which for a particle near the peak of the
valence shell density could be expected to be fairly deep and isotropic,
has been reduced to a shallow and open-ended dip.
As expected from
the form of the nondynamic correlation, little change was observed in the
PCF\ for a particle fixed on the $x$~axis.

A quantitative comparison of $g$ for the above
cases is also instructive and is shown in Figs.~\ref{atomfig3}(c) and (d).
We plot $g({\bf r_0},\uparrow;{\bf r},\downarrow)$ 
varying $r$ along the $x$~axis and fixing ${\bf r}_0$ at 1.4~$a_B$
on the $x$~axis (c) and the analogous situation for the $z$-axis (d).
Error bars on the statistical
measurements of these quantities are plotted, as well as the plane-wave cutoff
dependence for the $x$-axis case.  The converged result on the $x$~axis
(solid line) is well localized about the reference particle's 
position, with an enhancement of particle density on the opposite side 
of the atom.  

The results for the $z$~axis, Fig.~\ref{atomfig3}(d) 
include the PCF\ for the three different trial wavefunctions discussed above:
the single Slater determinant plus Jastrow factor,
$\psi_{S-J}$ (dashed line), the two-configuration wavefunction $\psi_{CI}$ 
(dotted line) and the same multiplied by a Jastrow factor $\psi_{CI-J}$ 
(solid line).
In comparison to the most accurate result,
$\psi_{CI-J}$, $\psi_{CI}$ predicts with some success
the long-wavelength polarization response to the electron at ${\bf r_0}$, 
determining how much the electron density is pushed from one side of the 
atom to the other.  The
major difference is the absence of the small dip in the immediate vicinity
of the electron due to the cusp condition, which naturally is not 
obtained from the two-configuration wavefunction.  
In contrast, $\psi_{S-J}$, which does not include explicit
orbital-dependent correlations but can be optimized to obtain
accurate results for short interparticle distances, is
nearly identical to $\psi_{CI-J}$ in the vicinity of the reference electron but
underestimates the long-wavelength polarization of the atom, obtaining
about $70\%$ of the PCF\ of $\psi_{CI-J}$ on the other side of the atom.
The mixing coefficient $\eta_1$ for the excited-state configuration
was reduced by about $60\%$ when
the Jastrow factor was added to the multiconfiguration result, 
indicating that part of the polarization of the atom produced by the
simple two configuration wavefunction is already accounted for by
the Jastrow factor.  In each case the $z$-axis PCF\ leads to a much
larger peak on the far side of the atom than the $x$-axis case.
 
\subsection{Error Analysis}
There are three sources of error in our calculation: statistical error
in taking Monte Carlo estimates, the finite plane-wave cutoff of the data,
and finally the variational bias due to the discrepancies of the trial
from the true ground state wavefunction.  The first two are closely connected
and to some extent can be regulated; the third is harder to assess.

In a typical Monte Carlo calculation, sample points for evaluating
the density or other single-particle expectation in a given region of space
are generated with frequency proportional to the density itself.
This leads to statistically precise measurements of the density at high
density and large relative errors that vary roughly as $1/\sqrt{n(r)}$
at the vanishingly low densities outside the atom.
Calculating the VMC expectations, $<\sum_i exp( -i{\bf G}\cdot {\bf r}_i )>$, 
of a set of plane-waves periodic on
a supercell is equivalent to taking the Fourier transform of a histogram 
distribution of statistical sample points on that cell. 
The statistical outliers from the poorly sampled, asymptotic low-density region
show up in the Fourier transform as a noise background independent of energy.
This noise can to some
degree be controlled by imposing a finite cutoff in reciprocal space.  
However, too small a cutoff leads to spurious long wavelength 
oscillations and is particularly poor for the short-distance region of the
hole where the cusp condition results in a long range tail in the reciprocal
space.  We find that a good balance between controlling statistical 
and plane-wave error can be found by choosing a cutoff when the statistical
average of the plane-wave component of the density is roughly equal to 
the statistical noise in its calculation.  For a sampling size of around
$10^5$ independent configurations, this cutoff limit proves to be about
32~Ry for a resolution of 1.0~$a_B$.  

The statistical error of the
Monte Carlo sampling was measured both for the individual plane-wave
components of the density and conditional density, and for their real-space
counterparts by measuring the standard deviation of these quantities over
10 to 20 independent runs.
As shown in Fig.~\ref{atomfig3}(c) and (d), the error bars of the PCF\ do
in fact vary roughly as $1/\sqrt{n(r)}$ with well controlled errors in the
peak density region, increasing to arbitrarily large values 
as one moves outside the atom.  With the cutoff used, statistical errors
in the PCF are limited to a value of less than
0.05 (out of a range in the PCF\ of the order of 1.0) for 
particles within 3.5~$a_B$ 
of the atom center, which in effect provides the limits in the plots of the 
PCF\ in Fig.~\ref{atomfig4}.

The convergence of the plane-wave expansion was checked by 
plotting the PCF\ as a function of cutoff energy.
A typical result is shown in Fig.~\ref{atomfig3}(c) 
where the $x$-axis PCF\ for a particle fixed at 
1.4~$a_B$ on the axis is plotted, for cutoff energies of 20, 28 and 32~Ry.
The exact on-top hole has been measured by a direct VMC calculation and is 
plotted as a circle.  
The 20~Ry calculation shows clear deviations
from the higher energy cutoff data, particularly in the region of the on-top
hole, ie., in the immediate vicinity of the fixed electron, where
the cusp condition contributes a high-energy tail to the exchange-correlation 
hole.  
The agreement between the two higher energy plots is 
well within statistical error except 
in the core region of the atom and in the low density tails.  
The 32~Ry case has not yet 
converged in the on-top region to the exact on-top value plotted as a circle,
indicative of the slow convergence of the plane-wave expansion to the 
on-top hole cusp.

A final source of error is from the effect of the deviation
of the variational trial 
wavefunctions~[Eqs.~(\ref{eqhfpsi}) and (\ref{eqcipsi})] 
from the exact ground state.  The ground-state energy, being variationally
optimized, is typically determined with much less error than other 
expectations (though one may expect that those important in the determination
of the energy, such as the density and $n_{xc}$, should still be robust.
An example of this problem is demonstrated in Fig.~\ref{atomfig4} in which 
two variationally optimized wavefunctions having relatively
insignificant differences in total energy, give PCF's with
noticeable differences for an electron fixed on the unoccupied $z$~axis.
The variational method is in general more sensitive to errors
in the wavefunction at high density as this region contributes the most
to the variational energy.  Thus we expect worse errors in the PCF\
when one or both coordinates are at low density.    

In order to gauge possible errors arising
from variational bias in our wavefunction, we study the change in the PCF\ 
and in $n_{xc}$ for several choices of trial wavefunction.  Specifically
we used trial wavefunctions with four, ten and eighteen 
Boys and Handy basis functions, corresponding
to keeping expansion terms of order $o=l+m+n$ up to 2, 4 and 6 
respectively in our 
Jastrow function.  The lowest order function has, in addition to two
terms that set the cusp condition for each spin component, only one
electron-electron and one electron-nucleus term.  The correlation energy
of each wavefunction is 0.0640(8), 0.0812(4), and 0.0840(2)~a.u. respectively,
in comparison to 0.0877(2)~a.u. using the multideterminant Slater-Jastrow
wavefunction and 0.0883~a.u. for the CI calculation of Ref.~\cite{Mitas}.

\begin{figure}[htb]
\epsfbox{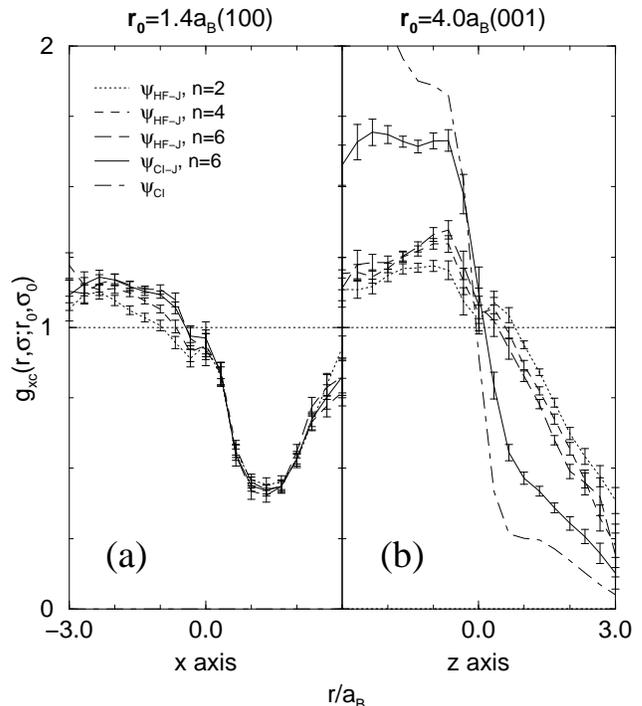}
\caption
{Convergence data for the opposite-spin PCF\ for
up-spin electron fixed at (a) 1.4~$a_B$ on the $x$~axis and (b) 4.0~$a_B$
on the $z$~axis.  Cases shown are for Boys and Handy terms up to order
$o=2,4, \text{and}\ 6$ as described in the text, in a one-determinant 
Slater-Jastrow 
wavefunction, and $o=6$ for a two-determinant Slater-Jastrow wavefunction.
Additionally, results for a two-determinant
wavefunction without Jastrow factor are given in (b) as a dot-dashed line.}
\label{atomfig5}
\end{figure}

In Fig.~\ref{atomfig5} we plot the opposite-spin PCF\ using these trial
wavefunctions for two cases previously considered in 
Sec.~\ref{results_nxc}: (a) an up-spin electron placed at 1.4~$a_B$ on the
$x$~axis and (b) an up-spin electron placed at 4.0~$a_B$ on the $z$~axis.
As in Fig.~\ref{atomfig3} we present a cut through the atom center and 
the reference electron, thus showing the minimum and maximum of the PCF.
The first case represents a probable best-case situation, with the reference
electron placed at peak density on an axis occupied by a $3p$ orbital,
so that nondynamic correlations due to the open-shell structure 
should be neglible.  In contrast the second
case presents the worst case scenario: at low density along the unoccupied
axis.  The behavior of the PCF\ in the asymptotic low-density region of the 
atom is dominated by the statistical error of our plane-wave basis 
and is not shown.

In our best-case scenario, (a), the PCF\ near the reference electron is
very well obtained with even the crudest model used ($o=2$).  
On the far side of the atom with respect to the reference electron position, 
there is a noticeable increase in the magnitude of the peak of the PCF\ 
with the increase in accuracy of the wavefunction.  As expected, the 
addition of nondynamic correlations in form of a two-determinant 
Slater-Jastrow wavefunction (solid line) has no noticeable effect.
An investigation of the corresponding contour plots, cutting through the
$x$-$z$ plane as shown in Fig.~\ref{atomfig4}, shows that the depth 
and spatial extent of the PCF\ is obtained with the lowest order
wavefunction and varies only slightly with increase in basis set.  
The major difference
is the gradual adjustment of the $g=1$ contour from a more isotropic shape
to the elliptical one shown in Fig.~\ref{atomfig4}(a), which accounts 
for the gradual increase in the peak on the far side of the $x$~axis as 
the $g=1$ contour moves towards the atom center.  

The worst case scenario (b) shows a far greater degree of disagreement between
wavefunctions.  The three one-determinant Slater-Jastrow wavefunctions 
agree fairly well with each other, with an increasing
percentage of the electron density removed from
the near side of the atom and placed on the far side.  The peak height 
on the far side is roughly the same as for the 1.4~$a_B$, high-density
case shown in Fig.~\ref{atomfig3}(d).  
The introduction of the nondynamic $3s^2\!\rightarrow\!3p^2$
excitation into the Slater-Jastrow wavefunction (solid line) leads to a 
dramatic change in the shape of the PCF.
The peak of the PCF\ from the base value of $g=1$ representing no
change due to correlations is increased by a factor of three, 
and only at the on-top hole (not shown) is there no significant change 
in the PCF.  
In comparison, the PCF\ of the optimized two-determinant
wavefunction $\psi_{CI}$ discussed in Sec.~\ref{results_pcf} is shown as 
a dot-dashed line.  The two-determinant Slater-Jastrow wavefunction
is an interpolation between the two limiting cases, favoring $\psi_{CI}$
at most locations.  

An explanation of the ``gigantic" features in the nondynamic part of the
PCF\ in Fig.~\ref{atomfig5}(b) comes from the
observation that the ratio $zR_p(r) / R_s(r)$ 
between the $3p$ and
$3s$ orbital, which determines the shape of the nondynamic PCF\ to first 
order~[Eq.~\ref{eqgci}],
increases exponentially at large distances along the $z$~axis.  As a result, 
in the asymptotic region (that is, either the response to a
reference electron outside the atom or the asymptotic tails of the response
to a reference electron at high density), the nondynamic correlation introduced
with the $3s^2\!\rightarrow\!3p_z^2$ substitution is no longer a small 
perturbation even with a small mixing parameter.  
The Jastrow factor basis functions however are chosen to 
tend to a constant at either large electron-electron
or electron-nucleus distances. 
It is quite likely that none of the cases studied accurately 
represents the true asymptotic behavior of the correlation function. 
They should rather be considered to provide a qualitative idea of the PCF\ 
as well as a sense of the range of behavior it should reasonably lie within.

It is interesting to note that the agreement between the various PCF's is far 
closer for the high density case on the $z$~axis plotted 
in Fig.~\ref{atomfig3}(d).
This indicates the greater weight of the exchange-correlation hole
at high density in determining the total exchange-correlation energy, 
and the corresponding robustness of its determination even
with qualitatively different wavefunctions.  
Given the large degree
of variation in the asymptotic limit, it would be interesting
to study the effect of including a larger number of configurations in 
a multideterminantal Slater-Jastrow wavefunction or include the
effects of backflow or multielectron coordinates~\cite{SM} into the
evaluation of orbitals.
At higher densities we expect the effect of such improvements in the 
wavefunction would be to add further refinements in 
the shape of the PCF\ along the $z$~axis, but with
much less change in basic features such as its range and magnitude.

\section{Correlation effects on the spin density}
\label{results_spin} 

In addition to the exchange-correlation hole, the spin components of the
single-particle density
change with the inclusion of Coulomb correlations.
In principle, the Kohn-Sham equations used to derive the density in DFT
should give the exact ground-state radial density, even for a degenerate
ground-state~\cite{Savin}; however there is no such principle for the 
spin components of the density.  Thus a change observed in the spin components
of the density that does not alter the total radial density can be considered 
an intrinsic feature of Coulomb correlation and not simply due to the 
inaccuracy of the LDA density.
As with $n_{xc}$, a prominent feature of 
the change in the spin-dependent density, $\Delta n({\bf r},\sigma)$, 
is anisotropy with respect to the fully occupied $x$ and open $z$ axes. 
This anisotropy in $\Delta n({\bf r},\sigma)$
may help to shed light on the mechanisms underlying the 
anisotropy in $n_{xc}$.

\begin{figure}[htb]
\epsfbox{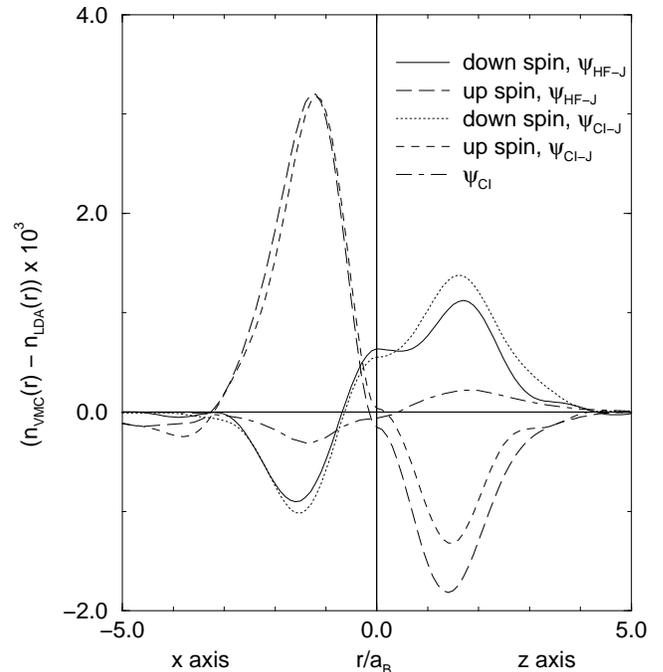}
\caption
{The~change~in~the~spin-dependent~density 
$n_1({\bf r},\sigma)\!-\!n_0({\bf r},\sigma)$ between 
the fully interacting and noninteracting systems.  The left side
of the plot shows a cut through the $x$~axis; the right side, 
through the $z$~axis.  The solid line is the spin-down
density change using the one-determinant Slater-Jastrow wavefunction 
$\psi_{S-J}$;
the long-dashed is the spin-up density change for the same wavefunction.  
Dotted and dashed lines
give the corresponding quantities for the two-determinant Slater-Jastrow
function $\psi_{CI-J}$.  The dot-dashed line is for the two-determinant 
function $\psi_{CI}$.  The plot units are $10^{-3}$~a.u. }
\label{atomfig6}
\end{figure}

In Fig.~\ref{atomfig3}, the spin-down density for the single determinant
Slater-Jastrow wavefunction is plotted as a dotted line 
along the $x$~axis (a) and $z$~axis (b).  Note that in the noninteracting
wavefunction this density is that of the $3s$ orbital and thus the
same along each axis.  However, in the interacting case, anisotropy in
the correlation response reduces the density along the $x$~axis by
roughly 6\% and increases it by the same amount
along the $z$-axis.  This change is shown in Fig.~\ref{atomfig6} (solid line)
along with the change in the spin-up density (long-dashed line), with the 
right side of the plot showing a cut through the $x$~axis and the left side
a cut through the $z$-axis.  The
corresponding density changes using the two-determinant Slater-Jastrow 
wavefunction ($\psi_{CI-J}$) are plotted as dotted and dashed lines.
In both cases, the change in the spin-up density has a qualitative trend 
opposite to that of the spin-down density, showing a large enhancement in 
density near the peak of the $3p_x$ orbital and a drop in density 
along the $z$~axis.  Thus, the changes in the two spin components
cancel out partially in the total density, but add to the total spin
density, defined as $m({\bf r}) = n({\bf r},\uparrow)-n({\bf r},\downarrow)$.
In addition there are observable
effects of a small reduction in the radius of the atom that occurs with the
addition of the Jastrow correlation term, particularly a reduction in 
up-spin density at a large distance from the atom and an enhancement of the 
down-spin density in the core region.  A convergence test along the lines
of that for $n_{xc}$ in the previous section was carried out for the 
spin components of the density.  The qualitative trends are repeatable with
the less accurate Jastrow factors but with considerably slower 
quantitative convergence than for $n_{xc}$.

An anisotropic shift in spin density can be generated from the
nondynamic Fermi correlation discussed in the previous sections:
the mixing of the noninteracting ground-state with
a $3s^2\! \rightarrow\! 3p_z^2$-substitution excited state  
and the lack thereof due to Pauli exclusion for the $3p_x$ and $3p_y$ analogs.  
Using the two-determinant wavefunction $\psi_{CI}$, the resulting change in each
spin component of the density is
\begin{equation}
      \Delta n({\bf r},\uparrow) = \Delta n({\bf r},\downarrow) =
	      \eta_1^2 (|\psi_{3p_z}({\bf r})|^2 - |\psi_{3s}({\bf r})|^2)
\end{equation}
where $\psi_{3s}$ and $\psi_{3p_z}$ are the wavefunctions of 
the $3s$ and $3p_z$ orbitals and
$\eta_1$ the excited state probability amplitude.
This function, shown as a dash-dotted line in Fig~\ref{atomfig6}, 
corresponds to a density enhancement along the $z$~axis and reduction
in the $x$-$y$ plane for {\it both} spin-components for a net zero 
change in the spin density $m({\bf r})$, in marked contrast to
the mutually opposing changes of the two spin-components of the Monte
Carlo data.  Given the optimal value of $\eta_1^2 = 0.017$ for the occupation
number of the excited-state configuration, one finds that the magnitude of
the change in either spin component of the density is much smaller than
observed, and qualitatively in the wrong direction for the up-spin case.
When the nondynamic $s^2 \rightarrow p_z^2$ excitation is included 
into the Slater-Jastrow wavefunction (dotted and dashed lines), both the
up-spin and down-spin densities are enhanced slightly along the $z$~axis,
but the qualitative picture remains unchanged -- as one might expect from
the very small value of $\eta_1^2=0.003$ that was optimal for this case.
For these reasons, it is unlikely that the observed anisotropic 
change in the spin components of the density can be explained by this type 
of mechanism.

On the other hand, a simple explanation of these results lies in that 
the observed change in the spin density reduces the spatial overlap between the
two spin components.  
As the correlation energy is predominantly determined by the spatial 
correlation between opposite-spin electrons, such a change in spin density
can reduce the correlation energy in a way that reduces the total energy 
if the total density remains unchanged.  
This correlation effect is similar to that of the unrestricted Hartree-Fock
method~\cite{Fulde} in which the energy of a system like ${\rm H_2}$
can be lowered from its Hartree-Fock value by breaking a symmetry of the 
ground state to induce the spatial separation of opposite spin-components
of the density.
In the Si atom, since the ground state is degenerate and lacks the symmetry
of the Hamiltonian, it already has a nonzero spin density; multiplication
by a Jastrow factor does not break the symmetry of the ground
state in a substantial way~\cite{HFUm}.  
The change in spin-density reflects
the effect of the Coulomb interaction which induces the spatial separation
between opposite spin electrons, in a system in which the different spin
components are to some degree already spatially separated in the 
noninteracting state.  In contrast to a filled-shell atom 
or other spin-unpolarized
system, this spatial separation of opposite spins appears not only as a
correlation hole but as a change in the mean density as well.
Consistent with this picture, the direction of the change in the $L_z\!=\!0$ 
projection of the Si atom enhances the absolute difference in the spin density 
$|n({\bf r},\uparrow)-n({\bf r},\downarrow)|$
that already exists in the Hartree-Fock
wavefunction, where up-spin electrons occupies $p_x$ and $p_y$ orbitals
while the lone down-spin electron does not.

\section{Summary and Conclusion}
\label{conclusion}
We have calculated 
the exchange-correlation hole of the valence shell of the ground state of the
Si atom as a function of spin decomposition, using the Variational Monte
Carlo method.  
This relatively simple four electron system, restricted to a single valence
shell, nevertheless shows a rich variety of 
phenomena in exchange and correlation not present in closed-shell atoms.

The incomplete filling of the valence shell of the open-shell atom leads 
to dramatic anisotropy and nonlocality in the exchange hole, even at peak 
densities in the valence shell.  This is accompanied by a significant
compensating anisotropy in the correlation hole making
the total exchange-correlation hole more (but not completely)
local and isotropic.  

Our paper has focused mostly on the exchange-correlation hole about a
majority (up) spin electron.  In this case, as one goes from a reference point
at peak density along an ``occupied" axis, along which one of the occupied
$3p$ orbitals is oriented, to one on the ``open" $z$~axis perpendicular 
to the $3p$ orbitals, the exchange hole changes from an efficiently screening 
$3sp^2$-like to a poorly screening $3s$ character.  As a result, we observe  
in the response of the minority (down) spin density
a significant ``dipole" shift 
or shift of density from one side of the atom to the other, that occurs 
when the up-spin electron is placed in the peak density position 
on the open axis and not when it is 
placed in the center of a $p$ orbital.  This difference
shows up in the pair correlation function as a difference in shape, with
that on the ``occupied" axis being a modest distortion of the isotropic shape of
a dynamic correlation, and that on the $z$~axis showing a marked dipole
component along the $z$~axis.  
The dipole polarization observed is also
notable in that it occurs not only for a reference electron outside
the atom but at the peak along the $z$~axis, where the local gradient is zero;
it is therefore a truly nonlocal feature not amenable to modeling by an 
expansion in the local density gradient.

The explicit inclusion of nondynamic correlations into our wavefunction
enhances the difference between open and occupied-axis response, 
particularly when the reference electron is moved outside the atom.  This
is due to the exclusion effect in which the $3s^2 \rightarrow 3p^2$ 
substitution is allowed along the open axis and excluded along the occupied
ones.  Nevertheless, the existence of such a difference in the 
single-determinant Slater-Jastrow wavefunction, given the structure of the
Jastrow factor, seems better explained in terms of the screening of the
exchange hole, in which the Coulomb interaction of the down spin electron
with the up-spin $3s$ exchange hole causes the dipole response along the
open axis.  In this case, it is the lowlying $3p$ orbital that best
compensates for the poor screening of the exchange hole.  It should thus 
play an important if not exclusive role in inducing the anisotropy 
of the opposite spin correlation hole.

A second major effect of the open-shell structure of the atom upon its
response to the Coulomb interaction is the change 
of the spin components of the density with the addition of 
correlations.  The
down-spin density is pushed inwards and onto the unoccupied axis and the
up-spin density pushed off the $z$~axis and onto the peak 
of the $p$ orbitals in the $x$-$y$ plane, resulting  
in a reduction of the spatial overlap between the two spin components
in a way which leaves the radial density largely 
unchanged.  The standard exclusion effect resulting
from the $3s^2\!\rightarrow\!3p_z^2$ substitution 
produces a net change in density that is both quantitatively too small
and qualitatively incorrect.  However, 
both the anisotropic features of the exchange-correlation hole and the
changes in spin density essentially stem from the same effect: the tendency
of the Coulomb interaction to induce the spatial separation of opposite
spin electrons in the context of the spatially anisotropic and 
spin-polarized structure of the degenerate Si ground state.

We have only briefly discussed the $n_{xc}$ for a down-spin electron, where
in addition to the screening of the exchange hole (always $3s$-like),
the anisotropy of the determinantal structure of the up-spin 
electrons plays an important role.  The correlation response to the 
down-spin electron on the occupied axis combines a compensation 
for poor screening by the exchange hole and the absence of a
lowlying $3p$-orbital component from the correlation
hole on account of Pauli exclusion.
We have observed in this situation subtle structural properties in the PCF\ 
that merit further investigation.

In contrast to longer ranged features of the correlation hole
we find that its on-top value -- the reduction of electron density
in the immediate vicinity of an electron -- 
is reproduced by a local density ansatz within 10-20\%, over
a fairly wide range in density and magnetization, and is otherwise insensitive
to the structure of the atom.  Given the energetically reasonable shape of 
the LDA $n_{xc}$ 
and that it satisfies the short-range cusp condition and global
particle-sum rule of the true $n_{xc}$ in addition to the approximate fit
to the Si on-top hole, it is reasonable to expect that
it will provide a good approximation for the angle- or system-averaged
$n_{xc}$,
that is, after averaging out many of the subtle angle or position dependent 
features studied here~\cite{EPB}. 
Nevertheless, the complex spin-dependent phenomena observed in this paper point to the inherent difficulty of systematically improving on local
or semilocal density functional theories in systems with nontrivial
valence-shell structure, such as open shell atoms or multiply-bonded
molecules.  

Our results provide support for several recent hybrid approaches to DFT.
The combination of a short ranged on-top hole that
is fairly well modeled by local density functional theory 
and longer-ranged exchange and exclusion
effects that are not indicates the
usefulness of density functional schemes which include the on-top
hole as a basic component~\cite{PEBS} or involve the
hybridization of 
short-ranged local or semilocal density functionals with a more accurate
treatment of longer ranged correlations using RPA~\cite{Langreth} or
CI~\cite{Savin}. 
The tendency of the correlation hole to cancel out the anisotropy in 
the exchange hole, and increase the sensitivity of the exchange-correlation
hole to electron position, 
along with the quality of the on-top hole in the LSD approximation,
lends support to recent hybrid approaches~\cite{Becke} 
which mix the exact exchange hole with an local DFT approximation of the 
full-coupling constant exchange-correlation hole.  These methods rely 
on the assumption that the full coupling constant limit of the integral
in Eq.~(\ref{eqexc}) is more likely to be amenable to approximation by 
the isotropic, localized $n_{xc}$ of the homogeneous electron gas than the
noninteracting limit dominated by exchange.
In the case of the open-shell atom
exclusion effects quite effectively cancel out the gross anisotropy in the
exchange hole caused by the open-shell structure, at least within the
region of peak valence density. 

On the other hand, the changes in spin-density that we observe
indicate that a standard Hartree-Fock or restricted CI basis set
may be a less desirable starting point
for implementing hybrid DFT methods in spin-polarized systems
than, for example, a generalized Hartree-Fock approach that would
allow for anisotropic distortions in the spin density due to Coulomb 
correlations.
Also, it seems possible that more could be done to obtain accurate correlation 
energies within the local spin-density approximation with the incorporation
of projection specific information.

One of us (A. C. Cancio) would like to thank Kieron Burke for helpful
discussions.
This work was supported by Sandia National Laboratories contract AP-7094
and the Campus Laboratories Collaboration of the University of California.


\begin{thebibliography}{99}
\bibitem{Coulomb} The exchange hole of DFT is nearly
identical with the Fermi hole commonly studied in quantum
chemistry.  However, the noninteracting ground-state in DFT and the
corresponding single-particle orbitals are determined by
the Kohn-Sham equation and 
in principle obtain the density of the exact ground state, while
the Fermi hole is usually defined in terms of the Hartree-Fock ground state.
Similarly, the correlation hole defined with respect to the full coupling
constant exchange-correlation hole is the analog of the Coulomb hole of
Ref.~\cite{Coulson}.
\bibitem{Node}
In principle one should
be able to measure the importance of this configuration by a calculation of 
the overlap between it and the Slater-Jastrow ground state;
the significantly different nodal structure of the two states makes a VMC
calculation of this quantity unreliable in practice.
\end{thebibliography}
\end{document}